\documentclass{article}
\usepackage{arxiv}
\usepackage[utf8]{inputenc} 
\usepackage[T1]{fontenc}    
\usepackage{hyperref}       
\usepackage{url}            
\usepackage{booktabs}       
\usepackage{amsfonts}       
\usepackage{nicefrac}       
\usepackage{microtype}      
\usepackage{lipsum}         
\usepackage{graphicx}
\usepackage{natbib}
\usepackage{doi}
\usepackage{subcaption}     
\usepackage{float}          
\usepackage{pdflscape}      
\usepackage{multirow}
\usepackage{amssymb}
\usepackage{enumitem}
\usepackage{afterpage}
\usepackage{caption}
\usepackage{listings}
\usepackage{lineno}
\usepackage{xcolor}

\definecolor{lightgray}{rgb}{0.9, 0.9, 0.9}

\lstdefinestyle{pseudocode}{
    xleftmargin=10mm,
    backgroundcolor=\color{lightgray},
    basicstyle=\small\ttfamily,
    numbers=left,
    numberstyle=\tiny\color{gray},
    commentstyle=\color{blue},
    keywordstyle=\color{purple},
    stringstyle=\color{olive},
    columns=flexible,
    mathescape=true,
    showstringspaces=false,
    breaklines=true,
    breakatwhitespace=true,
    tabsize=2
}

\lstnewenvironment{pseudocode}{\lstset{style=pseudocode}}{}

\title{Evaluation of Task Specific Productivity Improvements Using a Generative Artificial Intelligence Personal Assistant Tool}

\date{} 

\author{Brian S.~Freeman\thanks{Corresponding author: \texttt{brian.freeman@tranetechnologies.com}} \\
    Trane Technologies \\
    \And
    Kendall Arriola \\
    Trane Technologies \\
    \And
    Dan Cottell \\
    Trane Technologies \\
    \And
    Emmett Lawlor \\
    Trane Technologies \\
    \And
    Matt Erdman \\
    Trane Technologies \\
    \And
    Trevor Sutherland \\
    Trane Technologies \\
    \And
    Brian Wells \\
    Trane Technologies
}

\renewcommand{\shorttitle}{Productivity Improvements Using GenAI}

\hypersetup{
    pdftitle={Evaluation of Task Specific Productivity Improvements Using a Generative Artificial Intelligence Personal Assistant Tool},
    pdfsubject={q-bio.NC, q-bio.QM},
    pdfauthor={Brian S. Freeman, Kendall Arriola, Dan Cottell, Emmett Lawlor, Matt Erdman, Trevor Sutherland, Brian Wells},
    pdfkeywords={Generative AI, Productivity, Employee Assistant, Task Analysis, AI Integration}
}

\begin{document}
\maketitle

\begin{abstract}
This study evaluates the productivity improvements achieved using a generative artificial intelligence personal assistant tool (PAT) developed by Trane Technologies. The PAT, based on OpenAI's GPT 3.5 model, was deployed on Microsoft Azure to ensure secure access and protection of intellectual property. To assess the tool's productivity effectiveness, an experiment was conducted comparing the completion times and content quality of four common office tasks: writing an email, summarizing an article, creating instructions for a simple task, and preparing a presentation outline. Sixty-three (63) participants were randomly divided into a test group using the PAT and a control group performing the tasks manually. Results indicated significant productivity enhancements, particularly for tasks involving summarization and instruction creation, with improvements ranging from 3.3\% to 69\%. The study further analyzed factors such as the age of users, response word counts, and quality of responses, revealing that the PAT users generated more verbose and higher-quality content.
\begin{itemize}
    \item Write an email: 3.3\% improvement
    \item Summarize text: 69\% improvement
    \item Create instructions: 45.9\% improvement
    \item Prepare an outline: 24.8\% improvement
\end{itemize}
An '\textit{LLM-as-a-judge}' method employing GPT-4 was used to grade the quality of responses, which effectively distinguished between high and low-quality outputs. The findings underscore the potential of PATs in enhancing workplace productivity and highlight areas for further research and optimization.
\end{abstract}


\section{Introduction}
Trane Technologies developed a personal assistant tool (PAT) in the autumn of 2023 using a version of OpenAI's ChatGPT 3.5-Turbo-16k on Microsoft Azure cloud platform. Over the course of several weeks, the platform was introduced throughout all business units and international offices of the company and made available to all of its 40,000+ staff. 

\vspace{1em} 
\noindent To evaluate the improvement in productivity this tool provides, an experiment was initiated that compared the completion time and content of four common office tasks between a test group of users who could use the PAT and a control group of users who completed the tasks manually. The tasks included: 

\vspace{1em} 

\begin{itemize}
\item Writing an email to a supervisor
\item Summarizing an article
\item Creating instructions for a simple task
\item Preparing a presentation outline
\end{itemize}

\vspace{1em} 

\section{Background}

While large language models (LLMs) have been around since 2017 \citep{Vaswani2017}, their capabilities were not widely adopted until November 2022 when OpenAI announced the release of ChatGPT 3 \citep{openai2022}. Using generative pre-trained (GPT) models as personal assistants for text-based tasks became widespread, until several leaks of intellectual property were reported in the media and information on how the LLMs were trained using scraped content from the internet forced many companies to block access to Gen AI websites and services.

\noindent Realizing the need to allow controlled and secure access to GenAI tools, Trane Technologies developed an in-house tool based on OpenAI's GPT 3.5-Turbo-16k-0613 model but deployed on a Microsoft Azure platform that established usage guardrails and secured company intellectual property. Other companies followed suit, such as Proctor \& Gamble with their \textit{chatPG} \citep{PG2024}.
\vspace{1em} 

\noindent In order to quantify productivity improvement using PAT's, estimates of time savings are applied to usage metrics across the board, without consideration of the individual tasks. Estimations were based on published reports suggesting up to 54\% efficiencies for certain office tasks could be achieved using a PAT \citep{Noy2023}.
\vspace{1em} 

\noindent The Harvard Business School polled 758 business consultants with access to ChatGPT-4 with 18 different tasks \citep{Dell2023}. They found that use of AI allowed workers to complete 12.2\% more tasks 25.1\% faster with 40\% better quality, with less experienced workers improving even more than experienced workers. The tasks were specific to business consulting such as marketing, branding, and product management. The study introduced the idea of retainment - how much of the content generated by the AI was included in the final submitted answer. The degree of retainment was linked to how users applied the AI. The two categories of use were defined as \textit{Centaur} and \textit{Cyborg} behavior. Centaur behavior is inspired by the legendary creature that combines human and horse traits. It involves a close collaboration between humans and machines, with users alternating between AI and human tasks based on their strengths and capabilities. They assess which tasks are best suited for humans and which can be effectively managed by AI. Cyborg behavior was inspired by science fiction hybrids and combines machine components with human biology for seamless integration. Users go beyond a simple division of labor, intertwining their efforts with AI at the cutting edge of capabilities. Both behaviors impact how individuals utilize AI and PATs.

\noindent The National Bureau of Economic Research reported improvements of 14\% by customer service agents with higher increases by less-experienced workers compared to more-experienced workers \citep{Bryn2023}. One explanation of this enhancement is the reduction of required working memory load to handle large amounts of data or work under stressful conditions \citep{nielson2023}.
\vspace{1em} 

\noindent An MIT report published in the journal Science sampled 444 college educated professionals with specific writing tasks such as preparing a press release, writing a short report, analyzing a plan, and writing a delicate email \citep{Noy2023}. The participants' objective was to improve grades on their assignments and incentivized with cash rewards based on the points received. In addition to improving grade quality by 27\%, average task time was improved by 54\%. Noy et al. found that higher quality results improved at the cost of time needed to complete the task, and deduced that participants were spending the time editing the gen AI responses or modifying prompts to enhance responses. A key observation was that human input added value to the responses, even if it increased the overall time it took to complete the task. An interesting component about their methodology was that after users were given training on the AI and completed the first task, they were allowed to choose whether they wanted to use the AI for additional tasks. Follow-up questions were also provided that helped explain why gen AI PATs might not be used for some tasks. Respondents explained that the technology did not work well with specialized tasks or tasks that required up to date information to complete.

\section{Methods}

A web-based survey was prepared that included an introduction page that collected metadata on the respondent. After this data was collected the survey began. The time a respondent was on each page was measured and stored. 
\vspace{1em} 

\noindent Two versions of the survey were prepared. The control version blocked the ability of the respondent to paste content from the operating system's clipboard (No Paste), while the test version allowed respondents to paste from the clipboard. This allowed test group users to access the PAT and copy the results from their prompts and paste into the survey collection window. Control group users had to type their responses into the survey collection windows. Control group users could theoretically access the PAT and use it to generate a response, however no instance of this was identified.
\vspace{1em} 

\noindent The surveys were written in ReactJS and Tailwind, with the results stored in a Postgress database. Results were downloaded into a comma separated value (csv) file for analysis.  Respondents were randomly given one of the two URLs from 15-30 April 2024. Initially, the respondents were limited to individuals attending an on-site event at the corporate headquarters in Davidson, NC. Survey participants used their own laptops in reserved rooms with a monitor to explain how to access the survey and respond to specific questions. Later, URLs were emailed to volunteers in the company who were not in attendance. The respondents could ask questions through email and chat services.

\section{Survey Content}
Tasks were selected that were considered common for all users and did not need specific skills or subject matter expertise. These included: 

\vspace{1em} 

\begin{itemize}
\item Task 1: Writing an email to a supervisor (Email)
\item Task 2: Summarizing an article (Summary)
\item Task 3: Creating instructions for a simple task (Instructions)
\item Task 4: Preparing a presentation outline (Outline)
\end{itemize}

\vspace{1em} 

\noindent The actual tasks are provided in Appendix A. 

\section{Excluded Tasks}

Tasks that included specific knowledge such as coding or third party tools such as translations were not considered due to access to limited sub population groups within the company. Current use of the company PAT suggests that over 55\% of all prompts in the company are code related. Further research in this area as it related to company requirements should be considered. Current studies suggest coding tasks can be completed 56\% faster using gen AI resources \citep{Peng2023}, while another study found that over 52\% of PAT or Co-pilot assisted code contained errors \citep{kabir2024}. 

\section{Metadata}

Basic metadata was collected on all respondents as shown in Table \ref{tab:metadata}. All results were anonymous.

\begin{table}[]
\centering
\caption{Required metadata and selection options.}
\label{tab:metadata}
\begin{tabular}{@{}cc@{}}
\toprule
\textbf{Parameter} & \textbf{Options} \\ \midrule
Age &  \\
Gender &  \\
Education & HS/Bachelor/Masters/PhD \\
Company Position &  \\
Location &  \\
Work Location & Remote/Hybrid/On-Site \\
Experience with AI & None/Some/Beginner/Expert \\ \bottomrule
\end{tabular}
\end{table}

\vspace{1em} 

\noindent While most of the metadata parameters are evident, the \textit{Experience with AI} parameter was broken down as follows:
\vspace{1em} 

\begin{itemize}
\item None - No previous experience using the PAT
\item Some - Some experience using the PAT, but not weekly
\item Beginner - Regular use (at least weekly) of a PAT, but only the web interface
\item Expert - Regular use of a PAT and gen AI model using API interfaces
\end{itemize}

\vspace{1em} 
\noindent Respondents readily identified themselves with a category without additional prompting.
\clearpage

\section{Results}

Sixty-three (63) respondents provided input to this study. Thirty-two (32) were part of the Test Group (could paste from clipboard) and thirty-one (31) were part of the Control Group and could not paste. Their breakdowns are shown below.

\subsection{Metadata Results}

\begin{figure}
  \centering
  \begin{subfigure}{0.45\textwidth}
    \centering
    \includegraphics[width=\textwidth]{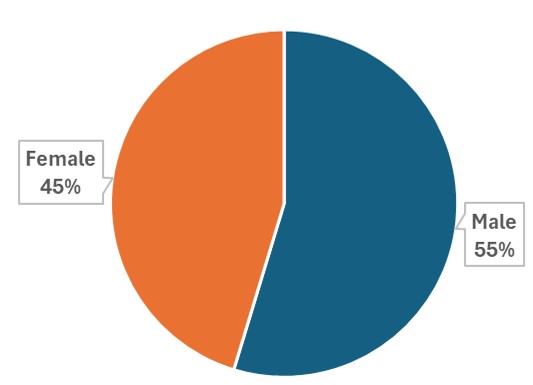}
    \caption{Gender}
  \end{subfigure}
  \hfill
  \begin{subfigure}{0.45\textwidth}
    \centering
    \includegraphics[width=\textwidth]{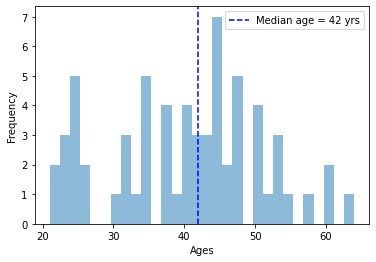}
    \caption{Age}
  \end{subfigure}
  \caption{Respondent gender and age}
  \label{fig:gender-age}
\end{figure}

There were slightly more male respondents than females with a median age of 42 years as shown in Figure \ref{fig:gender-age}.

\begin{figure}
  \centering
  \begin{subfigure}{0.45\textwidth}
    \centering
    \includegraphics[width=\textwidth]{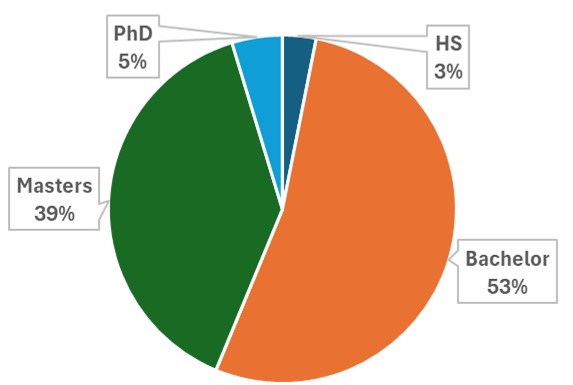}
    \caption{Education levels}
  \end{subfigure}
  \hfill
  \begin{subfigure}{0.45\textwidth}
    \centering
    \includegraphics[width=\textwidth]{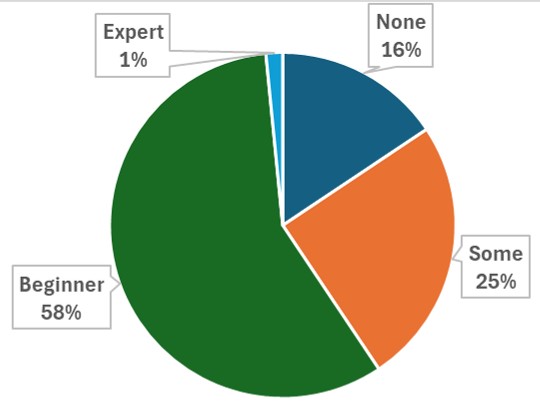}
    \caption{Experience with Generative AI}
  \end{subfigure}
  \caption{Respondent education and experience with gen AI}
  \label{fig:education-exp}
\end{figure}

\noindent Almost all of the respondents had college degrees, with almost half of them having graduate degrees. This might be attributed to the specialized population participating at the internal company data conference where the survey was given.  This also could explain the large number of individuals identifying as Beginner level (having regular exposure to the PAT prior to taking the survey) as shown in Figure \ref{fig:education-exp}.

\noindent Job positions were manually entered by the respondent resulting in 51 different position names. In order to classify each position into more manageable and comparable categories, the list of position titles was categorized by the PAT using the prompt: 

\begin{pseudocode}
User message = "Given this list of position titles, please group into no more than 5 different categories: $\ll$ list of position names $\gg$"
\end{pseudocode}

\noindent The resulting categories were:

\vspace{1em} 

\begin{itemize}
\item Quality Management System (QMS)
\item Data and Analytics
\item IT and Software Development
\item Finance and Pricing
\item Project and Operations Management
\end{itemize}

\vspace{1em} 

\begin{figure}
\centering
\includegraphics[width=0.9\textwidth]{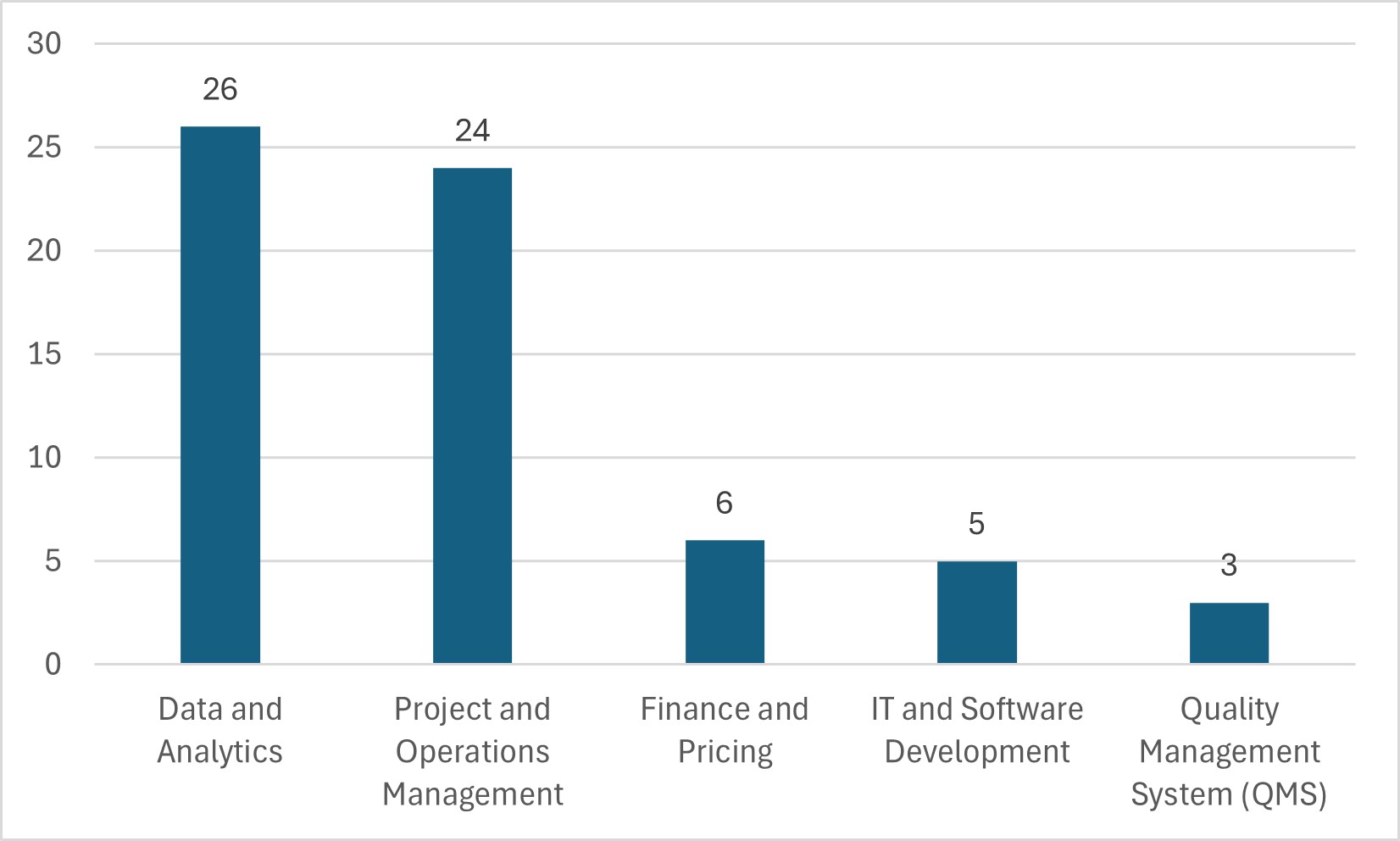}
\caption{Respondent roles}
\label{fig:role}
\end{figure}

\noindent Each category was assigned to the position titles given by the respondents. Figure \ref{fig:role} shows that the majority of participants were in Data \& Analytics or Project Management positions.  Seven of the respondents were interns or junior staff in the Accelerated Development Program (ADP) representing new employees.

\clearpage 

\subsection{Task Results}

\begin{figure}
  \centering

  \begin{subfigure}{0.45\textwidth}
    \centering
    \includegraphics[width=\textwidth]{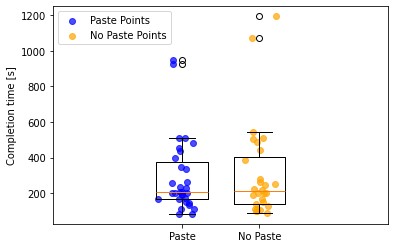}
    \caption{Task 1 - Email}
    \label{fig:sub1}
  \end{subfigure}
  \hfill
  \begin{subfigure}{0.45\textwidth}
    \centering
    \includegraphics[width=\textwidth]{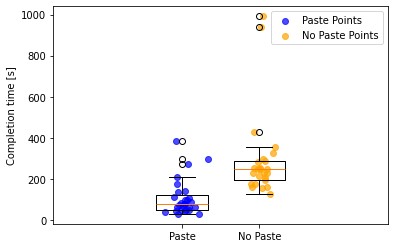}
    \caption{Task 2 - Summary}
    \label{fig:sub2}
  \end{subfigure}

  \vskip\baselineskip

  \begin{subfigure}{0.45\textwidth}
    \centering
    \includegraphics[width=\textwidth]{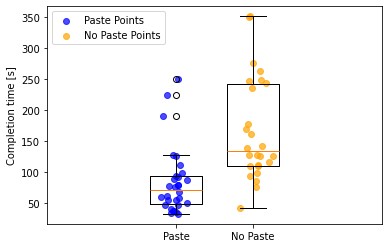}
    \caption{Task 3 - Instruction}
    \label{fig:sub3}
  \end{subfigure}
  \hfill
  \begin{subfigure}{0.45\textwidth}
    \centering
    \includegraphics[width=\textwidth]{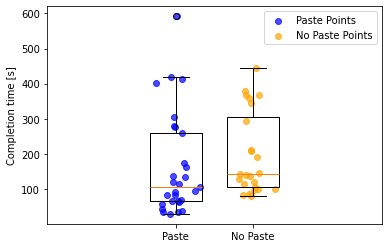}
    \caption{Task 4 - Outline}
    \label{fig:sub4}
  \end{subfigure}

  \caption{Aggregated results for tasks}
  \label{fig:overalltasks}
\end{figure}

The aggregated results are shown in Figure \ref{fig:overalltasks} for each task. The figures include all datasets without segregation by subgroups.

\noindent A review of the distributions of the aggregated data in each task suggested that all subsets were heavily skewed and not normally distributed. This was further confirmed by the normalcy test shown in Appendix B. The median of each subset was used for descriptive purposes and shown in Table \ref{tab:task_summary}.

\begin{table}
\centering
\caption{Median results for each task}
\label{tab:task_summary}
\begin{tabular}{@{}cccc@{}}
\toprule
\textbf{Task} & \textbf{No Paste [s]} & \textbf{Paste [s]} & \textbf{Improvement} \\ \midrule
\textbf{Email} & 211 & 204 & -3.3\% \\
\textbf{Summary} & 248 & 77 & -69.0\% \\
\textbf{Instruction} & 133 & 72 & -45.9\% \\
\textbf{Outline} & 141 & 106 & -24.8\% \\ \bottomrule
\end{tabular}
\end{table}

\noindent Reviewing the Improvement results in Table \ref{tab:task_summary} shows a wide variation of productivity enhancements, from 3.3\% for email generation to 69\% for summarization of text. 

\subsubsection{Statistical Evaluation of Aggregated Results}
To determine statistical significance of the aggregated results, a non-parametric test was used to account for the non-normal distribution of the data. In this case the Mann-Whitney U test (also known as the Wilcoxon rank sum test) was used \citep{mcknight2010}. The null hypothesis (H$_{O}$) and alternative hypothesis (H$_{1}$) were defined as:

\vspace{1em} 
\begin{itemize}
\item H$_{O}$: There is no difference between the Control Group and the Test Group.
\item H$_{1}$: There is a significant difference between the Control Group and the Test Group.
\end{itemize}

\vspace{1em} 

\noindent with a level of significance of p = 0.05 (5\%. The results of the test are shown in Table \ref{tab:results_mann}.

\begin{table}
\centering
\caption{Mann-Whitney U Test for significance (p = 0.05)}
\label{tab:results_mann}
\begin{tabular}{@{}cccc@{}}
\toprule
\textbf{Task} & \textbf{U Statistic} & \textbf{p-value} & \textbf{Result} \\ \midrule
\textbf{Email} & 426.5 & 0.915 & Fail to reject (Not Significant - Strong) \\
\textbf{Summary} & 626.5 & 8.73E-07 & Reject (Significant - Strong) \\
\textbf{Instruction} & 658.5 & 1.06E-05 & Reject (Significant - Strong) \\
\textbf{Outline} & 453 & 0.062 & Fail to Reject (Not Significant - Weak) \\ \bottomrule
\end{tabular}
\end{table}

\noindent Table \ref{tab:results_mann} shows that the Email and Outline tasks do not pass and the null hypothesis should not be rejected (no significant differences between the groups). However, the p-value for the Outline task is very close to the critical value of 0.05 and could be assumed to be significant (reject null). Both the Summary and Instruction tasks have strong differences from the null hypothesis and can safely be considered significantly different. The interquartile range shown in the box-plots of Figure \ref{fig:overalltasks} are shown in Table \ref{tab:task_ci} for 25\% and 75\% percentiles.

\begin{table}
\centering
\caption{Lower (25\%) and upper (75\%) quartiles in Figure \ref{fig:overalltasks}.}
\label{tab:task_ci}
\begin{tabular}{@{}ccccc@{}}
\toprule
\textbf{} & \multicolumn{2}{c}{\textbf{No Paste [s]}} & \multicolumn{2}{c}{\textbf{Paste [s]}} \\ 
\textbf{Task} & \textbf{25\% percentile} & \textbf{75\% percentile} & \textbf{25\% percentile} & \textbf{75\% percentile} \\ \midrule
\textbf{Email} & 87 & 542 & 83 & 511 \\
\textbf{Summary} & 128 & 355 & 29 & 210 \\
\textbf{Instruction} & 41 & 352 & 32 & 128 \\
\textbf{Outline} &80 & 445 & 29 & 419 \\ \bottomrule
\end{tabular}
\end{table}

\subsubsection{Evaluation of Age vs Completion Time}

\begin{figure}
  \centering

  \begin{subfigure}{0.45\textwidth}
    \centering
    \includegraphics[width=\textwidth]{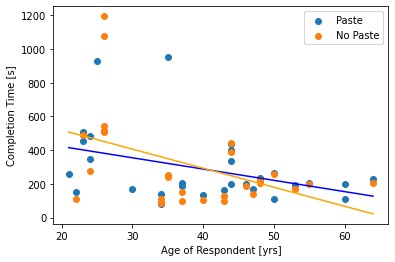}
    \caption{Task 1 - Email}
    \label{fig:agesub1}
  \end{subfigure}
  \hfill
  \begin{subfigure}{0.45\textwidth}
    \centering
    \includegraphics[width=\textwidth]{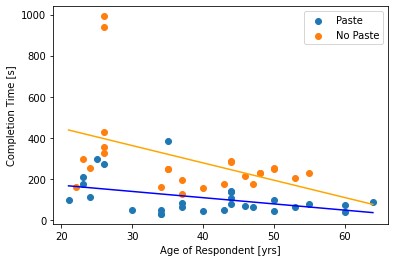}
    \caption{Task 2 - Summary}
    \label{fig:agesub2}
  \end{subfigure}

  \vskip\baselineskip

  \begin{subfigure}{0.45\textwidth}
    \centering
    \includegraphics[width=\textwidth]{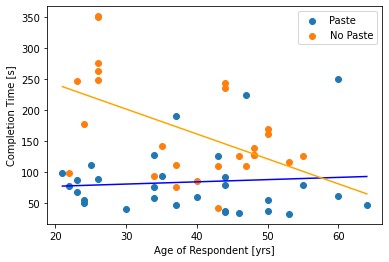}
    \caption{Task 3 - Instruction}
    \label{fig:agesub3}
  \end{subfigure}
  \hfill
  \begin{subfigure}{0.45\textwidth}
    \centering
    \includegraphics[width=\textwidth]{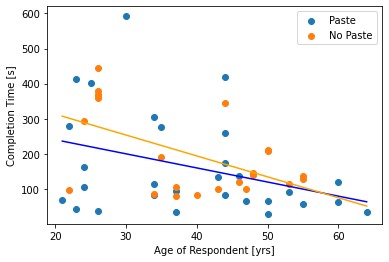}
    \caption{Task 4 - Outline}
    \label{fig:agesub4}
  \end{subfigure}

  \caption{Evaluation of respondent's age vs completion time for tasks}
  \label{fig:age-tasks}
\end{figure}

Determining how different users utilized the PAT was an important objective of the study. Physical age was used to indirectly represent user experience with the general assumption that the participant had similar experience throughout their individual careers \citep{morris2006}. Figure \ref{fig:age-tasks} shows scatter plots of completion time to complete tasks by age of the participants with simple overlayed linear regression trend lines. The slopes and intercepts of the trend lines are given in Table \ref{tab:age-time}.

\noindent Spearman's Rank Correlation or Spearman's $\rho$ \citep{spearman1904} was used to show correlation between respondent age and completion time. This statistic is a non-parametric measure of correlation and commonly used when analyzing the relationship between two variables that may not follow a normal distribution and is shown along with the trend line coefficients in Table \ref{tab:age-time}.

\begin{table}
\centering
\captionsetup{width=.75\textwidth}
\caption{Coefficients and correlations for trend lines in Figure \ref{fig:age-tasks}}
\label{tab:age-time}
\begin{tabular}{@{}lcccccc@{}}
\toprule
\textbf{} & \multicolumn{3}{c}{\textbf{Paste}} & \multicolumn{3}{c}{\textbf{No Paste}} \\ 
\textbf{Task} & \textbf{Slope} & \textbf{Y-int} & \textbf{Spearman's $\rho$} & \textbf{Slope} & \textbf{Y-int} & \textbf{Spearman's $\rho$} \\ \midrule
\textbf{Email} & 0.26 & 212.26 & -0.086 & 3.24 & -41.66 & -0.182 \\
\textbf{Summary} & 0.06 & 101.31 & 0.113 & 8.42 & -100.41 & 0.181 \\
\textbf{Instructions} & -0.09 & 106.00 & 0.092 & 1.39 & 82.63 & -0.407 \\
\textbf{Outline} & 0.28 & 101.83 & 0.141 & 2.00 & 103.32 & -0.038 \\ \bottomrule
\end{tabular}
\end{table}

\vspace{1em} 

\noindent  Correlation is not strong, with the highest being a negative correlation for Task 3 (Instructions) for the No Paste scenario.

\vspace{1em} 
\noindent Overall trends suggest decreasing completion times the older and more experienced the user was. This was especially evident for the Control Group (No Paste) with all tasks experiencing negative slopes on linear regression trend lines except for Task 3. In this case, the Test Group (Paste)showed a slight positive trend based on respondent age. For all cases, the No Paste scenario had steeper rates then the Paste scenarios.

\vspace{1em} 

\begin{figure}
  \centering

  \begin{subfigure}{0.45\textwidth}
    \centering
    \includegraphics[width=\textwidth]{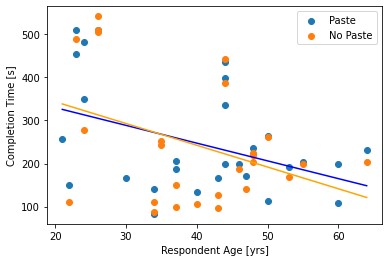}
    \caption{Task 1 - Email}
    \label{fig:zagesub1}
  \end{subfigure}
  \hfill
  \begin{subfigure}{0.45\textwidth}
    \centering
    \includegraphics[width=\textwidth]{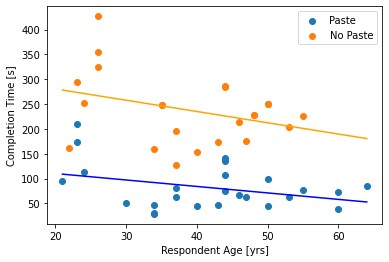}
    \caption{Task 2 - Summary}
    \label{fig:zagesub2}
  \end{subfigure}

  \vskip\baselineskip

  \begin{subfigure}{0.45\textwidth}
    \centering
    \includegraphics[width=\textwidth]{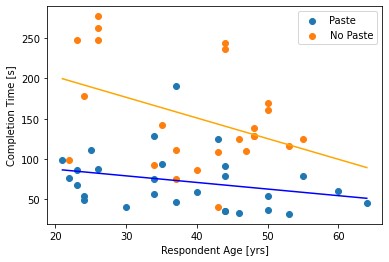}
    \caption{Task 3 - Instruction}
    \label{fig:zagesub3}
  \end{subfigure}
  \hfill
  \begin{subfigure}{0.45\textwidth}
    \centering
    \includegraphics[width=\textwidth]{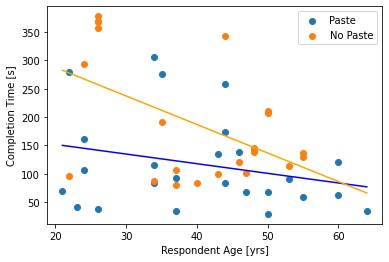}
    \caption{Task 4 - Outline}
    \label{fig:zagesub4}
  \end{subfigure}

  \caption{Evaluation of respondent's age vs completion time for tasks (filtered)}
  \label{fig:z-age-tasks}
\end{figure}

\noindent Concerns that some of the extreme points were influencing the linear regression lines led to evaluation of the main clusters for comparison. The Modified Z-Score test was used to identify outliers. Unlike the traditional Z-Score method, which assumes data follows a normal distribution, the Modified Z-Score method leverages the median and the Median Absolute Deviation (MAD), making it more resilient to the influence of outliers and suitable for non- normally distributed samples. The method calculates the modified z-scores for each data point by subtracting the median and then dividing by the MAD, scaled by a constant factor of 0.6745. This scaling factor ensures that the modified z-scores approximate standard z-scores for normally distributed data. Data points with an absolute modified z-score exceeding a chosen threshold of 3.5 for this case\citep{Hoaglin2013}. The filtered datasets are shown in Figure \ref{fig:z-age-tasks} with updated coefficients and Spearman's correlations in Table \ref{tab:z-age-time}.

\vspace{1em}

\vspace{1em} 
\noindent Filtering out extreme points shows more distinct clustering between the Paste and No Paste scenarios, especially with Task 2. Negative trends are present for all tasks and scenarios, with the No Paste slopes slightly steeper for all tasks compared to Paste scenarios.   One explanation is that older respondents are also more experienced and can accomplish these routine tasks manually faster then less experienced respondents. Filtering also improves correlation ranking in general, with strong negative correlation present in Task 3 for the Paste scenario.

\begin{table}
\centering
\captionsetup{width=.75\textwidth}
\caption{Coefficients and correlations for filtered trend lines in Figure \ref{fig:z-age-tasks}}
\label{tab:z-age-time}
\begin{tabular}{@{}lcccccc@{}}
\toprule
\textbf{} & \multicolumn{3}{c}{\textbf{Paste}} & \multicolumn{3}{c}{\textbf{No Paste}} \\ 
\textbf{Task} & \textbf{Slope} & \textbf{Y-int} & \textbf{Spearman's $\rho$} & \textbf{Slope} & \textbf{Y-int} & \textbf{Spearman's $\rho$} \\ \midrule
\textbf{Email} & -4.12 & 412.2 & -0.213 & -5.05 & 444.5 & -0.156 \\
\textbf{Summary} & -1.3 & 136.42 & -0.140 & -2.27 & 326 & -0.211 \\
\textbf{Instructions} & -0.82 & 103.7 & -0.910 & -2.56 & 253.4 & -0.156 \\
\textbf{Outline} & -1.7 & 185.8 & -0.298 & -5.04 & 388.4 & -0.175 \\ \bottomrule
\end{tabular}
\end{table}

\subsubsection{Evaluation of Response Word Counts}

\begin{figure}
  \centering

  \begin{subfigure}{0.45\textwidth}
    \centering
    \includegraphics[width=\textwidth]{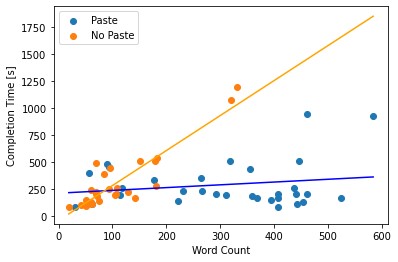}
    \caption{Task 1 - Email}
    \label{fig:wordsub1}
  \end{subfigure}
  \hfill
  \begin{subfigure}{0.45\textwidth}
    \centering
    \includegraphics[width=\textwidth]{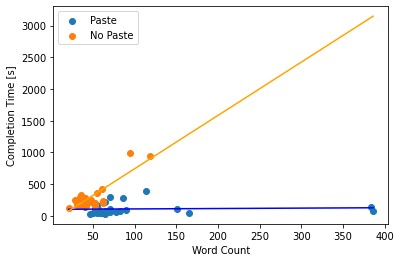}
    \caption{Task 2 - Summary}
    \label{fig:wordsub2}
  \end{subfigure}

  \vskip\baselineskip

  \begin{subfigure}{0.45\textwidth}
    \centering
    \includegraphics[width=\textwidth]{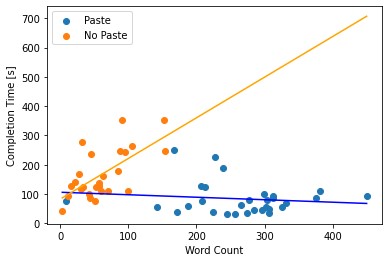}
    \caption{Task 3 - Instruction}
    \label{fig:wordsub3}
  \end{subfigure}
  \hfill
  \begin{subfigure}{0.45\textwidth}
    \centering
    \includegraphics[width=\textwidth]{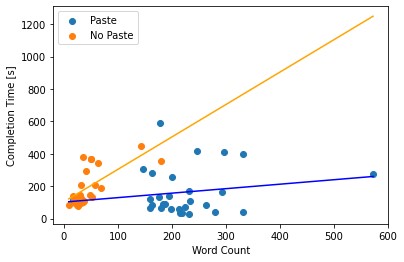}
    \caption{Task 4 - Outline}
    \label{fig:wordsub4}
  \end{subfigure}
  \caption{Evaluation of respondent's word count vs completion time for tasks}
  \label{fig:word-tasks}
\end{figure}

Another factor that was considered was how strong were the individual responses for both groups. In order to consider this, word counts for each task were evaluated. Word counts for each task and scenario are show in Figure \ref{fig:word-tasks} with coefficients of the trend lines in Table \ref{tab:word-tasks}.

\noindent Correlations in Table \ref{tab:word-tasks} were calculated using the same method as described in Table \ref{tab:age-time} with Spearman's $\rho$. Overall correlations are better than the previous section, especially for No Paste scenarios.

\begin{table}
\centering
\captionsetup{width=.75\textwidth}
\caption{Coefficients and correlations of trend lines in Figure \ref{fig:word-tasks}}
\label{tab:word-tasks}
\begin{tabular}{@{}ccccccc@{}}
\toprule
 & \multicolumn{3}{c}{\textbf{Paste}} & \multicolumn{3}{c}{\textbf{No Paste}} \\ 
\textbf{Task} & \textbf{Slope} & \textbf{Y-int} & \textbf{Spearman's $\rho$} & \textbf{Slope} & \textbf{Y-int} & \textbf{Spearman's $\rho$} \\ \midrule
Email & 0.26 & 212.26 & 0.029 & 3.24 & -41.66 & 0.813 \\
Summary & 0.06 & 101.31 & 0.307 & 8.42 & -100.41 & 0.352 \\
Instructions & -0.09 & 106.00 & 0.017 & 1.39 & 82.63 & 0.518 \\
Outline & 0.28 & 101.83 & -0.012 & 2.00 & 103.64 & 0.711 \\ \bottomrule
\end{tabular}
\end{table}

\noindent All tasks show similar patterns. The Paste scenario have more verbose answers with little fluctuation of completion time, while the No Paste scenario follow a positive correlation between word count and completion time with samples clustered around a central tendency. The median word counts for each task and scenario are shown in Table \ref{tab:word-medians}. A five-fold increase in text should be expected for most PATs, consistent with verbose reports in other studies \citep{Noy2023}.

\begin{table}
\centering
\captionsetup{width=.75\textwidth}
\caption{Median word counts for each task and scenario}
\label{tab:word-medians}
\begin{tabular}{@{}cccc@{}}
\toprule
\textbf{Task} & \textbf{No Paste} & \textbf{Paste} & \textbf{Increase} \\ \midrule
\textbf{Email} & 81 & 355 & 438\% \\
\textbf{Summary} & 41 & 66 & 161\% \\
\textbf{Instruction} & 55 & 276 & 502\% \\
\textbf{Outline} & 36 & 213 & 592\% \\ \bottomrule
\end{tabular}
\end{table}

\begin{figure}
  \centering

  \begin{subfigure}{0.45\textwidth}
    \centering
    \includegraphics[width=\textwidth]{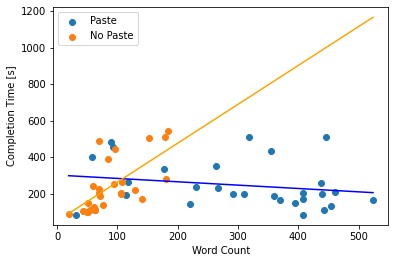}
    \caption{Task 1 - Email}
    \label{fig:zwordsub1}
  \end{subfigure}
  \hfill
  \begin{subfigure}{0.45\textwidth}
    \centering
    \includegraphics[width=\textwidth]{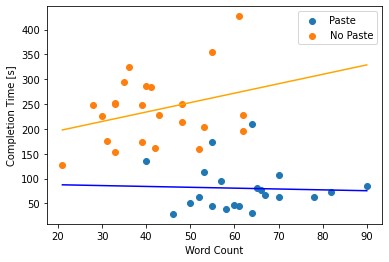}
    \caption{Task 2 - Summary}
    \label{fig:zwordsub2}
  \end{subfigure}

  \vskip\baselineskip

  \begin{subfigure}{0.45\textwidth}
    \centering
    \includegraphics[width=\textwidth]{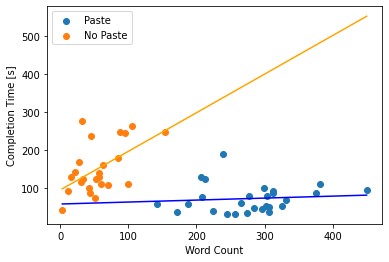}
    \caption{Task 3 - Instruction}
    \label{fig:zwordsub3}
  \end{subfigure}
  \hfill
  \begin{subfigure}{0.45\textwidth}
    \centering
    \includegraphics[width=\textwidth]{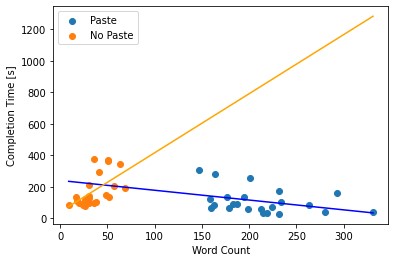}
    \caption{Task 4 - Outline}
    \label{fig:zwordsub4}
  \end{subfigure}
  \caption{Evaluation of respondent's word count vs completion time for tasks (filtered)}
  \label{fig:zword-tasks}
\end{figure}

\noindent As with concern of extreme values in the previous section comparing age to completion time, similar filtering was accomplished to evaluate the influence of outliers to the main clusters. A similar approach, using the Modified Z-Score Test,  was used. Results are shown in  Figure \ref{fig:zword-tasks} with coefficients of the trend lines in Table \ref{tab:zword-tasks}.

\noindent Figure \ref{fig:zword-tasks} shows more distinct clustering between the Paste/No Paste scenarios for each task, but reinforces the general patterns from Figure \ref{fig:word-tasks} in that the No Paste scenario tasks have a high word count rates but within a very short range (not exceeding 100 words for most tasks) while Paste scenarios generate more than 5x the words.

\begin{table}
\centering
\captionsetup{width=.75\textwidth}
\caption{Coefficients and correlations of filtered trend lines in Figure \ref{fig:word-tasks}}
\label{tab:zword-tasks}
\begin{tabular}{@{}ccccccc@{}}
\toprule
 & \multicolumn{3}{c}{\textbf{Paste}} & \multicolumn{3}{c}{\textbf{No Paste}} \\ 
\textbf{Task} & \textbf{Slope} & \textbf{Y-int} & \textbf{Spearman's $\rho$} & \textbf{Slope} & \textbf{Y-int} & \textbf{Spearman's $\rho$} \\ \midrule
\textbf{Email} & -0.18 & 301.4 & -0.175 & 2.13 & 50.1 & 0.766 \\
\textbf{Summary} & -0.17 & 91.2 & 0.120 & 1.9 & 158.2 & 0.160 \\
\textbf{Instructions} & 0.05 & 58.40 & 0.193 & 1.02 & 95.1 & 0.415 \\
\textbf{Outline} & -0.62 & 241 & -0.364 & 3.74 & 42.1 & 0.654 \\ \bottomrule
\end{tabular}
\end{table}

\subsubsection{Evaluation of Response Quality}

Multiple methods were provided by literature to evaluate response quality including Google's Universal Sentence Encoder \citep{cer2018} and LIWC \citep{boyd2022}. However, a zero-shot '\textit{LLM-as-a-judge}' method was used employing OpenAI's instance of GPT-4 on Microsoft Azure to act as a grader for the content. The GPT temperature was set to 0.0 to reduce hallucination effects and provide result reproducability \citep{zheng2023}. Each response, both No Paste and Paste scenario, was given to the GPT in the form of a prompt along with the task and a grading method. The prompt looked like this:

\begin{pseudocode}
System message = "You are an English language expert who will grade a response to a question.                
 Give a grade of 1 for a response that does not answer the question. 
 Give a grade of 2 for a response that answers the question with spelling and grammar errors. 
 Give a grade of 3 for a response that answers the question with good English. 
 Only provide a response that is 1, 2, or 3."

User message = "The question is: {task}. The user response is {response}"
\end{pseudocode}

\noindent The result was a numerical grade from 1 to 3 for each task response \citep{leng2023}. The review was run three times and an integer average of the three runs was used as the grade. Using a GPT-4 model to evaluate GPT-3.5 results from the PAT was assumed Evaluating GPT-3.5 results from the PAT using a GPT-4 model was considered appropriate due to the significant differences in training sets and methodologies between the two models, ensuring sufficient separation for objective evaluation.

\begin{figure}
  \centering

  \begin{subfigure}{0.45\textwidth}
    \centering
    \includegraphics[width=\textwidth]{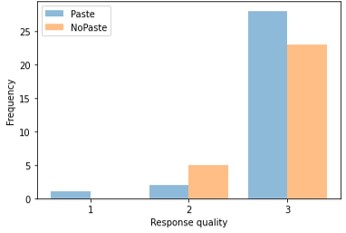}
    \caption{Task 1 - Email}
    \label{fig:qualsub1}
  \end{subfigure}
  \hfill
  \begin{subfigure}{0.45\textwidth}
    \centering
    \includegraphics[width=\textwidth]{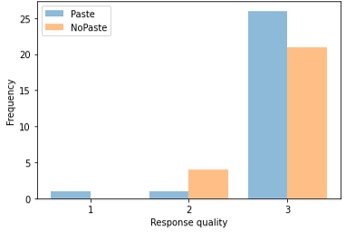}
    \caption{Task 2 - Summary}
    \label{fig:qualsub2}
  \end{subfigure}

  \vskip\baselineskip

  \begin{subfigure}{0.45\textwidth}
    \centering
    \includegraphics[width=\textwidth]{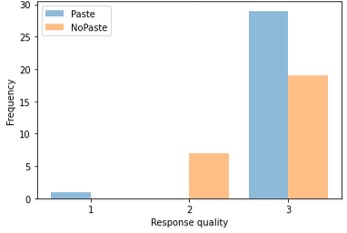}
    \caption{Task 3 - Instruction}
    \label{fig:qualsub3}
  \end{subfigure}
  \hfill
  \begin{subfigure}{0.45\textwidth}
    \centering
    \includegraphics[width=\textwidth]{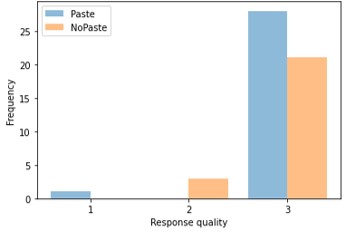}
    \caption{Task 4 - Outline}
    \label{fig:qualsub4}
  \end{subfigure}
  \caption{Evaluation of respondent's writing quality for tasks}
  \label{fig:qual-tasks}
\end{figure}

\noindent The GPT-4 reviewer correctly identified incomplete responses which were removed from the analysis pool. The remaining response samples were relatively close to each other, with predominately more lower-quality results from the No Paste respondents than the Paste group.

\clearpage

\section{Discussion}

Evaluation of the respondents' results looked at overall completion time to complete a task and then evaluating the task times against age, word count and quality of the responses. In general, the results reinforced observations from previous studies as well as highlighting some issues. Overall, using a PAT reduces the overall time to accomplish certain tasks while improving the quality of the output. Using PATs also generate responses that are more verbose.

\vspace{1em} 

\noindent Most interesting was the time required to write an email was almost the same between the Control and Test Groups. Although other studies acknowledged that users often spend time editing outputs or re-prompting to obtain better results, they did not evaluate how the final responses were achieved in terms of time spent and information retention.

\vspace{1em} 

\noindent Decomposing the different times in the total completion time for the Control Group that could not use the PAT (No Paste scenario) shows the following terms for $t_{no paste}$:

\begin{equation}
t_{no paste} = t_{ingest} + t_{transcribe} + t_{edit}
\label{eq:nopaste_full}
\end{equation}

\vspace{1em} 

\noindent where $t_{ingest}$ is the total time needed to read the task question and understand the requirements, $t_{transcribe}$ is the time needed to write the initial response in the content window, and $t_{edit}$ is the time used to edit the response. In many cases $t_{edit}$ = 0 when the the initial typed response is accepted. All terms are independent random variables.

\vspace{1em} 

\noindent For the Test group using the PAT, completion time, $t_{paste}$ terms becomes more complicated.

\begin{equation}
t_{paste} = t_{ingest} + 2t_{navigate} + \sum_{i = 1}^{n}\left ( t_{prompt_{i}} +  t_{edit_{i}} + t_{latency_{i}} + t_{response_{i}}\right ) + t_{copy} + t_{paste}
\label{eq:paste_full}
\end{equation}

\vspace{1em} 

\noindent where $2t_{navigate}$ is the time taken to navigate from the survey URL to the PAT URL, and then return to the survey URL after the PAT response has been copied. It can be assumed that the same time is required for both actions. 

\vspace{1em} 

\noindent For other terms, $t_{prompt}$ is the time used to write the prompt request for the PAT, $t_{latency}$ is the time the PAT uses to ingest the prompt, generate a response, and display the response, $t_{response}$ is the time to review the review the response, $t_{copy}$ is the time needed to copy the response from the PAT window, and $t_{paste}$ is the time needed to paste in the survey content window. The act of creating a prompt, editing it, waiting for the response and reviewing the response may be iterative in order to get a satisfactory result. To account for these iterations, $n$ represents the number of iterations of prompting used. As with \ref{eq:nopaste_full}, all the variables are independent and random.

\vspace{1em} 

\noindent Assuming $2t_{navigate}$, $t_{copy}$, and $t_{paste}$ are negligible, Eq \ref{eq:paste_full} can be reduced to 

\begin{equation}
t_{paste} = t_{ingest} + \sum_{i = 1}^{n}\left ( t_{prompt_{i}} +  t_{edit_{i}} + t_{latency_{i}} + t_{response_{i}}\right )
\label{eq:paste_reduced}
\end{equation}

\vspace{1em} 

\noindent Comparing the terms associated with Eq \ref{eq:nopaste_full} and \ref{eq:paste_reduced} it is clear that not using a PAT is less complicated in regards to user steps, especially for simple tasks.  More complicated tasks or tasks that may require more working memory load, such as summarizing a large body of text, will benefit from using a PAT, even if the text uses specialty language. 

\clearpage

\section{Conclusions and Recommendations}

A survey was prepared and presented to 63 employees of Trane Technologies to evaluate the productivity improvements of a generative AI base personal assistant tool (PAT). Half of the participants received a link to a survey where they could paste the results of the survey tasks from the Trane Technology PAT - Employee Virtual Assistant into the survey, while the other half could not paste content and had to complete the tasks manually. The resulting analysis showed that improvements varied extensively by task:

\vspace{1em} 

\begin{itemize}
\item Write an email: 3.3\% improvement using a PAT
\item Summarize text: 69\% improvement using a PAT
\item Create instructions: 45.9\% improvement using a PAT
\item Prepare an outline: 24.8\% improvement using a PAT
\end{itemize}

\vspace{1em} 

\noindent Further analysis showed that age of the user influenced overall time required to complete tasks, especially when summarizing a document without using a PAT. PATs were also shown to generate more verbose responses. This may ultimately result in longer emails, reports, and other text-based products, requiring more time to ingest by a user, or creating the need to use a PAT to complete more work tasks.

\vspace{1em} 

\noindent An '\textit{LLM-as-a-judge}' method using a GPT-4 model was used to review content quality that assigned a grade to each response based on the task and completion response. The method was able to filter out bad responses and assign grades to each response. Users who could not paste from the PAT had slightly lower quality than PAT users.

\vspace{1em} 

\noindent The small sample size of the study compared to other studies (63 participants compared to 444 in the lowest of other studies) may have biased results by not capturing enough variance in task completion time, word count, and content quality. Nonetheless, the diversity of participants that were collected suggest that results may not vary significantly, and at the very least, validate the expectation that different administrative tasks have different associated productivity efficiencies.  

\vspace{1em} 

\subsection{Recommendations}

\noindent This study looked at use cases at the nascent stage of PAT adoption. Evaluating productivity after a year of use where more people feel comfortable using the tools and have integrated them into work routines is recommended. Additionally, a similar study to see how well PATs impact coding and software development with light coders (data scientist, analysts, and engineers - users who use software and generate code but are not full time developers) should be conducted, especially since over 50\% of the queries received by company's PAT are code related.

\section{Supplementary information}

Raw survey datasets are available in csv format by contacting \url{brian.freeman@tranetechnologies.com}.

\section{Acknowledgements}
This project was an initiative executed under the Trane Technology AI Foundry. We wish to thank Rebecca Wells and Stephanie Benson for their support and encouragement. Also thanks to Phillip Sime for technical support and access to cloud infrastructure.

\bibliographystyle{unsrtnat}
\bibliography{task-study-bib}   

\vspace{1em} 
\Large
\textbf{Appendix}  
\renewcommand{\thesection}{\Alph{section}} 
\renewcommand{\thefigure}{\thesection.\arabic{figure}} 
\renewcommand{\thetable}{\thesection.\arabic{table}} 
  
\setcounter{figure}{0} 
\setcounter{table}{0} 
\setcounter{section}{0} 
  
\section{Survey Tasks}
The following four tasks were presented in a web-based survey. 

\subsection{Task 1 - Email}

Write an email to your supervisor requesting approval to go to the Data Analytics Summit. Include 3 reasons why he/she should approve it.

\subsection{Task 2 - Summary}

Summarize the following text: 

\begin{quote}
\setlength\leftskip{1cm} 
\setlength\rightskip{1cm} 
Generative Artificial Intelligence (AI) has revolutionized machine learning by enabling computers to exhibit creativity and generate content. The inception of generative AI can be traced back to the 1950s and 1960s when computer scientists began experimenting with automated systems capable of producing original content. The 1980s saw significant progress through the development of algorithms and computational techniques, resulting in the creation of systems capable of generating content in various domains.

\vspace{1em} 

The 1990s marked a significant milestone with the introduction of recurrent neural networks (RNNs) by researchers like Sepp Hochreiter and Jurgen Schmidhuber. RNNs allowed machines to generate sentences, scripts, and even poetry, resulting in increasing excitement in the field. However, it was the early 2000s that witnessed a remarkable resurgence in generative AI. This resurgence was fueled by the rise of deep learning and the introduction of generative adversarial networks (GANs) by Ian Goodfellow and his colleagues in 2014.

\vspace{1em} 

Deep learning algorithms utilizing neural networks with multiple layers greatly enhanced the generative capabilities of computers. GANs, in particular, revolutionized the field by introducing a generator-discriminator architecture, dramatically improving the quality and realism of generated content. This breakthrough enabled the generation of high-resolution images, music, and even human-like videos, opening up new possibilities and applications for generative AI.

\vspace{1em} 

Generative AI has found diverse applications across various industries. In creative fields such as art, design, and entertainment, it has pushed the boundaries of creativity and facilitated the generation of novel content. Industries like marketing and advertising have utilized generative AI to create personalized campaigns and generate content at scale, enhancing customer engagement.

\vspace{1em} 

Looking towards the future, researchers are continuously exploring methods to further enhance the diversity, coherence, and controllability of generated content. However, challenges, including ethical considerations related to the misuse of generative AI, need to be addressed. With ongoing innovations, the potential impact of generative AI on artistic expression, human-computer interaction, and society at large is vast, making it an exciting frontier in technology.
\end{quote}

\vspace{1em} 
\noindent Word count = 323 words

\subsection{Task 3 - Instructions}
Explain how to address a letter for postage.

\subsection{Task 4 - Outline}
Prepare an outline for a presentation that shows the major events from your last trip.

\clearpage 

\end{document}